# EDITORIAL: STATISTICS AND FORENSIC SCIENCE

By Stephen E. Fienberg

*Carnegie Mellon University*

Forensic science is usually taken to mean the application of a broad spectrum of scientific tools to answer questions of interest to the legal system. Despite such popular television series as *CSI*: *Crime Scene Investigation* and its spinoffs—*CSI*: *Miami* and *CSI*: *New York*—on which the forensic scientists use the latest high-tech scientific tools to identify the perpetrator of a crime and always in under an hour, forensic science is under assault, in the public media,[1] popular magazines [Talbot (2007), Toobin (2007)] and in the scientific literature [Kennedy (2003), Saks and Koehler (2005)]. Ironically, this growing controversy over forensic science has occurred precisely at the time that DNA evidence has become the "gold standard" in the courts, leading to the overturning of hundreds of convictions many of which were based on clearly less credible forensic evidence, including eyewitness testimony [Berger (2006)].

At the U.S. National Academies of Science/National Research Council, there have been symposia [Fienberg (2005)], reports [Committee to Review the Scientific Evidence on the Polygraph (2003), Committee on Scientific Assessment of Bullet Lead Elemental Composition Comparison (2004)] and other publications [Finneran (2003)] on various forensic scientific methods, all of which have raised serious questions about how virtually every form of forensic evidence except DNA comparisons has been used. Statisticians have played a prominent role in this ongoing debate over the uses and credibility of forensic science.

In this issue, Spiegelman et al. (2007) revisit the forensic evidence on the composition of bullet fragments found in 1963 at the scene of the assassination of President John F. Kennedy as well as the testimony about this evidence presented by a leading forensic scientist to the House Select Commitee on Assassinations. What is especially innovative in this article is not the methodology, which draws on the Committee on Scientific Assessment of Bullet Lead Elemental Composition Comparison (2004), but rather the fact that the authors were able to acquire a box of bullets from the same batch as bullets allegedly purchased by Lee Harvey Oswald, the putative single assassin, and then carry out compositional analyses of this new "sample" for comparison purposes. Their conclusions have stirred considerable public controversy and, even prior to the formal publication of the paper, they have been subject to extensive scrutiny.

[1]For example, see "CNN Presents Classroom: Reasonable Doubt: Can Crime Labs Be Trusted?" http://www.cnn.com/2005/EDUCATION/10/19/cnnpce.reasonable.doubt/index.html

The AOAS editors encourage our readers to judge for themselves the persuasiveness of this reassessment of the original bullet evidence and testimony.

Department of Statistics
 and Machine Learning Department
Carnegie Mellon University
Pittsburgh, Pennsylvania 15213
USA
E-mail: fienberg@stat.cmu.edu